
\newcount\notenumber

\def\note{\global\advance\notenumber by 1
\footnote{{\mathsurround=0pt$^{\the\notenumber}$}}
}

\mag=1300
\normalbaselineskip=11.66pt
\normallineskip=2pt minus 1pt
\normallineskiplimit=1pt
\normalbaselines
\vsize=17.3cm
\hsize=12.1cm
\parindent=20pt
\smallskipamount=3.6pt plus 1pt minus 9pt
\abovedisplayskip=1\normalbaselineskip plus 3pt minus 9pt
\belowdisplayskip=1\normalbaselineskip plus 3pt minus 9pt
\skip\footins=2\baselineskip
\advance\skip\footins by 3pt

\mathsurround 2pt
\newdimen\leftind \leftind=0cm
\newdimen\rightind \rightind=0.65cm

\def\pagenumbers{\footline={\hss\tenrm\folio\hss}}
\nopagenumbers




\font\sevenrm=cmr7
\font\seventeenbf=cmbx12 at 14pt

\def\MainHead{{\baselineskip 16.7pt\seventeenbf
\noindent\titolo\par}
\normalbaselines
\vskip 8.34pt
\noindent\autori\par
\ni\indirizzo
\footnote{\phantom{i}}{\piedipagina}
\vskip 23.34pt
\ni {\bf Abstract.} \Abstract
\np
}

\font\tenrm=cmr10
\font\tenit=cmti10
\font\tensl=cmsl10
\font\tenbf=cmbx10
\font\tentt=cmtt10

\def\tenpoint{%
\def\rm{\fam0\tenrm}%
\def\it{\fam\itfam\tenit}%
\def\sl{\fam\slfam\tensl}%
\def\tt{\fam\ttfam\tentt}%
\def\bf{\fam\bffam\tenbf}%
}

\tenpoint\rm

%
%
%
%
%
\font\ChapTitle=cmbx12
\font\SecTitle=cmbx12
\font\SubSecTitle=cmbx12
%
%
%
\def\NormalSkip{\parskip=5pt\parindent=15pt\baselineskip=12pt\PageNumbers%
\leftskip=0cm\rightskip=0cm}

\def\PageNumbers{\footline={\hss\tenrm\folio\hss}}
\def\NoPageNumbers{\footline={}}

%
%
%
\def\np{\vfill\eject}
\def\ss{\vskip 5pt}
\def\ms{\vskip 15pt}
\def\bs{\vskip 30pt}
\def\hs{\vskip 60pt}
\def\ni{\noindent}
%
%
%
\newcount\CHAPTER		          %
\newcount\SECTION		          %
\newcount\SUBSECTION		       %
\newcount\FNUMBER     
%
%
%
\CHAPTER=0		          
\SECTION=0		          
\SUBSECTION=0		       
\FNUMBER=0		          
%
%
%
\long\def\NewChapter#1{\global\advance\CHAPTER by 1%
\np\NoPageNumbers\ \vfil\parindent=0cm\leftskip=2cm\rightskip=2cm{\ChapTitle #1\hfil}%
\vskip 4cm\ \vfil\eject\SECTION=0\SUBSECTION=0\FNUMBER=0\NormalSkip}
%
%
%
\gdef\Mock{}
\long\def\NewSection#1{\bs\global\advance\SECTION by 1%
\ni{\SecTitle \ifnum\CHAPTER>0 \the\CHAPTER.\fi\the\SECTION.\ #1}\ms\SUBSECTION=0\FNUMBER=0}
\def\CurrentSection{\global\edef\Mock{\the\SECTION}} %
%
%
\long\def\NewSubSection#1{\global\advance\SUBSECTION by 1%
{\SubSecTitle \ifnum\CHAPTER>0 \the\CHAPTER.\fi\the\SECTION.\the\SUBSECTION.\ #1}\ss\FNUMBER=0}
%
%
%
%
\def\RightFormulaNumber{\global\advance\FNUMBER by 1\eqno{\fopen\the\FNUMBER\fclose}}
\def\LeftFormulaNumber{\global\advance\FNUMBER by 1\leqno{\fopen\the\FNUMBER\fclose}}
\def\RightFormulaLabel#1{\global\advance\FNUMBER by 1%
\eqno{\fopen\the\FNUMBER\fclose}\global\edef#1{\fopen\the\FNUMBER\fclose}}
\def\LeftFormulaLabel#1{\global\advance\FNUMBER by 1%
\leqno{\fopen\the\FNUMBER\fclose}\global\edef#1{\fopen\the\FNUMBER\fclose}}
%
%
%
\def\HeadNumber{\ifnum\CHAPTER>0 \the\CHAPTER.\fi%
\ifnum\SECTION>0 \the\SECTION.\ifnum\SUBSECTION>0 \the\SUBSECTION.\fi\fi}
\def\Getfn{\global\advance\FNUMBER by 1 {\fopen\HeadNumber\the\FNUMBER\fclose}}
\def\Getfl#1{\global\advance\FNUMBER by 1 {\fopen\HeadNumber\the\FNUMBER\fclose}%
\global\edef#1{\fopen\HeadNumber\the\FNUMBER\fclose}}
\def\ComposedRightFormulaNumber{\global\advance\FNUMBER by 1%
\eqno{\fopen\HeadNumber\the\FNUMBER\fclose}}
\def\ComposedLeftFormulaNumber{\global\advance\FNUMBER by 1%
\leqno{\fopen\HeadNumber\the\FNUMBER\fclose}}
\def\ComposedRightFormulaLabel#1{\global\advance\FNUMBER by 1%
\eqno{\fopen\HeadNumber\the\FNUMBER\fclose}%
\global\edef#1{\fopen\HeadNumber\the\FNUMBER\fclose}}
\def\ComposedLeftFormulaLabel#1{\global\advance\FNUMBER by 1%
\leqno{\fopen\HeadNumber\the\FNUMBER\fclose}%
\global\edef#1{\fopen\the\FNUMBER\fclose}}
%
%
%
\def\TheoremNumber{\global\advance\FNUMBER by 1 \fopen\the\FNUMBER\fclose}
\def\TheoremLabel#1{\global\advance\FNUMBER by 1\fopen\the\FNUMBER\fclose%
\global\edef#1{\fopen\the\FNUMBER\fclose}}
%
%
%
\def\ComposedTheoremNumber{\global\advance\FNUMBER by 1 \fopen\HeadNumber\the\FNUMBER\fclose}
\def\ComposedTheoremLabel#1{\global\advance\FNUMBER by 1\fopen\HeadNumber\the\FNUMBER\fclose%
\global\edef#1{\fopen\HeadNumber\the\FNUMBER\fclose}}
%
%
%
\def\fopen{(}\def\fclose{)}                 
\def\fn{\ComposedRightFormulaNumber}        
\def\fl{\ComposedRightFormulaLabel}         

\def\Compare#1#2{\message{^^J Compara \noexpand #1:=#1[#2]^^J}}

%
%
%
%

\def\ni{\noindent}
\def\ss{\vskip 5pt}
\def\ms{\vskip 10pt}

\def\noex{\noexpand}

\def\refs{}
\def\empty{\#}
\def\BibNumber{}
\def\BibTitle{}

\newcount\BNUM
\BNUM=0

\def\bib#1#2{\gdef#1{\global\def\BibNumber{\empty}\global\def\BibTitle{#2}}}

\def\ref#1{#1
\if\BibNumber\empty \global\advance\BNUM 1
\message{reference[\BibNumber]}\message{}
\global\edef\refs{\refs \ss\ni[\the\BNUM]\ \BibTitle}
\global\edef#1{\noex\global\noex\edef\noex\BibNumber{[\the\BNUM]}
 \noex\global\noex\edef\noex\BibTitle{\BibTitle}}
{\bf [\the\BNUM]}
\else
{\bf \BibNumber}
\fi}

\def\Biblio{{\refs}}


\def\SO{\hbox{\rm SO}}
\def\GL{\hbox{\rm GL}}

\def\Im{\hbox{\rm Im}}

\def\Diff{\hbox{\rm Diff}}

\def\Tr{\hbox{\rm Tr}}

\def\ds{\hbox{\bf ds}}
\def\d{\hbox{\rm d}}
\def\dt{\hbox{\rm dt}}
\def\dr{\hbox{\rm dr}}

\def\Aut{\hbox{\rm Aut}}
\def\dim{\hbox{\rm dim}}

\def\Lie{\hbox{\it \$}}

\def\calF{{\cal F}} 
\def\calL{{\cal L}}

\def\calU{{\cal U}} 
\def\calB{{\cal B}} 
 
\def\calS{{\cal S}}


\def\La{\Lambda}
\def\na{\nabla}
\def\la{\lambda}
\def\Si{\Sigma}
\def\si{\sigma}

\def\Om{\Omega}
\def\om{\omega}
\def\ep{\epsilon}
\def\al{\alpha}

\def\Ga{\Gamma}
\def\ga{\gamma}
\def\De{\Delta}
\def\de{\delta}

\def\te{\theta}

\def\R{{\Bbb R}}

\def\gBTZ{g_{_{\hbox{\sevenrm BTZ}}}}


\def\R{I \kern-.36em R}
\def\E{I \kern-.36em E}
\def\F{I \kern-.36em F}
\def\Co{I \kern-.66em C}
\def\id{1 \kern-.36em I}              

\def\del{\partial}                   

\def\QDE{{\offinterlineskip\lower1pt\hbox{\kern2pt\vrule width0.8pt
\vbox to8pt{\hbox to6pt{\leaders\hrule height0.8pt\hfill}\vfill%
\hbox to6pt{\hrulefill}}\vrule\kern3pt}}}


\def\arr{\rightarrow }            
\def\larr{\longrightarrow }            
\def\QDE{\hbox{\ }\vrule height4pt width4pt depth0pt}                                                              

\def\GL{\hbox{\rm GL}}

\def\np{\vfill\eject}
\def\ni{\noindent}

\def\ss{\vskip 5pt}
\def\ms{\vskip 10pt}
\def\bs{\vskip 15pt}
\def\hs{\vskip 20pt}

\bib{\WaldA}{V. Iyer and R. Wald, Phys. Rev. D {\bf 50},  1994, 846}

\bib{\WaldB}{R.M.\ Wald, J.\ Math.\ Phys., {\bf 31}, 1993, 2378 }

\bib{\WaldC}{B.\ S.\ Kay, R.\ M.\ Wald, Physics Reports {\bf 207}, (2),
North-Holland, 1991, 49}

\bib{\WaldD}{ R.\ M.\ Wald, Coordinate di Kruskal, ???}


\bib{\CADM}{M.\ Ferraris and M.\ Francaviglia, Atti Sem. Mat. Univ. Modena, {\bf 37},
1989, 61} 

\bib{\CADMB}{M.\ Ferraris, M.\ Francaviglia and I.\ Sinicco, Il Nuovo Cimento, {\bf 107B},
(11), 1992, 1303}

\bib{\CADMC}{M.\ Ferraris and M.\ Francaviglia, Gen.\ Rel.\ Grav., {\bf 22}, (9), 1990}

\bib{\Lagrange}{M.\ Ferraris, M.\ Francaviglia,
in: {\it Mechanics, Analysis and Geometry: 200 Years after Lagrange},
Editor: M. Francaviglia, Elsevier Science Publishers B.V., 1991}

\bib{\Cavalese}{M.\ Ferraris and M.\ Francaviglia, in: {\it 8th Italian
Conference on General Relativity and Gravitational Physics}, Cavalese (Trento), August 30 --
September 3, (World Scientific, Singapore, 1988) p. 183 }

\bib{\FFFG}{L.\ Fatibene, M.\ Ferraris and M.\ Francaviglia, M.\ Godina, Gen.\ Rel.\ Grav.\ {\bf
30} (9), 1371 (1998)}

\bib{\Brno}{L. Fatibene, M. Ferraris, M. Francaviglia, M. Godina,
{\it A geometric definition of Lie derivative for Spinor Fields},
in: Proceedings of 
{\it ``6th International Conference on Differential Geometry
and its Applications, August 28--September 1, 1995"}, (Brno, Czech Republic),
Editor: I. Kol{\'a}{\v r}, (MU University, Brno, Czech Republic, 1996) p. 549}

\bib{\Bologna}{L.\ Fatibene, M.\ Francaviglia, in; {\it Seminari di Geometria 1996-1997,
Universit\`a degli Studi di Bologna, Dipartimento di Matematica},  edited by S. Coen (Tecnoprint,
Bologna, 1998) p. 69}

\bib{\Robutti}{M.\ Ferraris, M.\ Francaviglia and O.\ Robutti, in:{\it G\'eom\'etrie et Physique},
Proceedings of the {\it Journ\'ees Relativistes 1985} (Marseille, 1985), 112 -- 125; Y.\
Choquet-Bruhat, B.\ Coll, R.\ Kerner, A.\ Lichnerowicz eds. Hermann, Paris, 1987}

\bib{\Kolar}{I.\ Kol{\'a}{\v r}, P.\ W.\ Michor, J.\ Slov{\'a}k, 
{\it Natural Operations in Differential Geometry} 
(Springer--Verlag, New York, 1993)}


\bib{\RevA}{A. Trautman, in: {\it Gravitation: An Introduction to Current Research}, L. Witten ed.
Weley, New York, 1962, 168}

\bib{\Katz}{J.\ Katz, Class.\ Quantum Grav., {\bf 2}, 1985, 423}

\bib{\BTZRef}{M.\ Ba\~nados, C. Teitelboim, J.\ Zanelli, Phys.\ Rev.\ Lett.\ {\bf 69}, 1849
(1992)}

\bib{\BTZRefB}{J.\ D.\ Brown, J.\ Creighton, R.\ B.\ Mann, Phys. Rev. D{\bf 50}, 6394 (1994)}

\bib{\BTZRefC}{D.\ Cangemi, M.\ Leblanc, R.\ B.\ Mann, Phys. Rev. D{\bf 48}, 3606 (1993)}

\bib{\BTZRefD}{S.\ Carlip, C.\ Teiteilboim, Phys.\ Rev.\ D{\bf 51} (2), 622 (1995)}

\bib{\BTZMann}{S.\ Carlip, J.\ Gegenberg, R.\ B.\ Mann, Phys.\ Rev.\ D{\bf 51} (12), 6854 (1995)}

\bib{\CaGe}{S.\ Carlip, J.\ Gegenberg, Phys.\ Rev.\ D{\bf 2} (44), 424 (1991)}

\bib{\Remarks}{L.\ Fatibene, M.\ Ferraris, M.\ Francaviglia, M.\ Raiteri, Ann.\ Phys.\ (in press);
e-archive: hep-th/9810039}

\bib{\Hawking}{ S.\ W.\ Hawking, C.\ J.\ Hunter, D.\ N.\ Page,
hep-th/9809035; S.\ W.\ Hawking, C.\ J.\ Hunter, hep-th/9808085;C.\
J.\ Hunter, hep-th/9807010}

\bib{\BrownYork}{J.\ D.\ Brown, J.\ W.\ York, Phys.\ Rev.\ D{\bf 47} (4), 1407 (1993)}

\bib{\TAUBNUT}{L.\ Fatibene, M.\ Ferraris, M.\ Francaviglia, M.\ Raiteri, {\it On the Entropy of
Taub-NUT Black Hole Solutions}, (in preparation)}

\bib{\DualA}{ M.\ Ferraris, M.\ Francaviglia, M.\ Raiteri, Quaderni del Dipartimento di Matematica
{\bf 35} (1998) (submitted to J. Math. Phys.)}

\bib{\DualB}{ R.\ Capovilla, T.\ Jacobson, J.\ Dell, Phys. Rev. Lett. {\bf 63}, 2325 (1989)}

\bib{\DualC}{ R.\ Capovilla, T.\ Jacobson, J.\ Dell, L.\ Mason, Class. Quantum. Grav. {\bf 8},
41 (1991)}

\bib{\DualD}{ R.\ Capovilla, T.\ Jacobson, J.\ Dell, Class. Quantum. Grav. {\bf 8}, 59 (1991)}

\bib{\DualE}{M.\ Raiteri, M.\ Ferraris, M.\ Francaviglia,  in: {\it ``Gravity, Particles and
Space-Time''}, edited by P.\ Pronin and G.\ Sardanashvily (Word Scientific, Singapore, 1996) p. 81}

\bib{\DualF}{G.\ Magnano, M.\ Ferraris, M.\ Francaviglia,  J.\ Math.\ Phys.\ {\bf 31} (2),
378 (1990)}

\bib{\XI}{M.\ Ferraris, J.\ Kijowski, Rend.\ Sem.\ Mat.\ Univers.\ Politecn.\ Torino {\bf 41},
3 (1983)}

\bib{\XII}{M.\ Ferraris, J.\ Kijowski, Gen.\ Rel.\ Grav.\ {\bf 14} (2), 165 (1982)}

\bib{\XIV}{F.\ A.\ Lunev, Phys.\ Lett.\ {\bf 295}B, 99 (1992)}

\bib{\XVI}{P.\  Peldan, Class.\ Quantum Grav.\ {\bf 9}, 2079 (1992)}

\bib{\Palatini}{M.\ Ferraris, J.\ Kijowski, Letters in Math. Phys. {\bf 5} 127 (1981); M.\ 
Ferraris, M.\  Francaviglia, C.\ Reina, Gen.\ Rel.\ Grav.\ {\bf 14} (3), 243 (1982); M.\ Ferraris,
J.\ Kijowski, Rend.\ Sem.\ Mat.\ Univers.\ Politecn.\ Torino, {\bf 41} (3), 169 (1983)}

\bib{\PartOne}{L.\ Fatibene, M.\  Ferraris, M.\  Francaviglia, M. Raiteri, 
{\it Remarks on Conserved Quantities and Entropy of BTZ Black Hole Solutions. Part I: The General
Setting}, (submitted)}

\bib{\Maeda}{J.-I.\ Koga, K.-I.\ Maeda, {\it
Equivalence of black hole thermodynamics between
a generalized theory of gravity and the Einsten theory},
e-print:gr-qc/9803086}



\def\titolo{Remarks on Conserved Quantities and Entropy of BTZ Black Hole Solutions. Part II: BCEA
Theory}
\def\autori{%
L. Fatibene, M. Ferraris, M. Francaviglia, M. Raiteri
}

\def\indirizzo{Dipartimento di Matematica, Universit\`a di Torino,\goodbreak
 via C. Alberto 10, 10123 Torino Italy}

\def\Abstract{
The BTZ black hole solution for $(2+1)$-spacetime is considered as a solution of a triad-affine
theory (BCEA) in which topological matter is introduced to replace the cosmological constant 
in the model.
Conserved quantities and entropy are calculated via N\"other theorem, reproducing in a geometrical
and global framework earlier results found in the literature using local formalisms.
Ambiguities in global definitions of conserved quantities are considered in detail.

A dual and covariant Legendre transformation is performed to re-formulate BCEA theory as a
purely metric (natural) theory (BCG) coupled to topological matter.
No ambiguities in the definition of mass and angular momentum arise in BCG theory.
Moreover, gravitational and matter contributions to conserved quantities and entropy are isolated.
Finally, a comparison of BCEA and BCG theories is carried out by relying on the results obtained in
both theories.
 }

\def\piedipagina{}

\vglue 33.3pt
\raggedbottom

\baselineskip=15pt

\MainHead

\np
\pagenumbers

\NewSection{Introduction}

This paper is the continuation of \ref{\PartOne}, which is hereafter referred to as Part I.
Notation follows closely.

In Part I we applied a geometrical, global and general formalism to study conserved quantities of
BTZ black hole solution (see \ref{\BTZRef}, \ref{\BTZRefB}, \ref{\BTZRefC}, \ref{\BTZRefD}, 
\ref{\BTZMann} and \ref{\CaGe}) in $(2+1)$ dimensions.
These particular solutions are realistic enough to be extensively used as test models for various
approaches to conserved quantities and entropy in General Relativity. We hereafter consider
so-called BCEA theory (see \ref{\CaGe}, \ref{\BTZMann}) which admits a BTZ black hole solution in a
frame-affine formalism, as well as the corresponding purely metric theory, which will be called BCG
theory. The BCG theory is obtained from BCEA theory by means of a ``dual'' covariant Legendre
transformation. This is an application of a fairly general framework which applies to a much
wider class of covariant theories (see for example \ref{\DualA}, \ref{\DualB}, \ref{\DualC},
\ref{\DualD}, \ref{\DualE}, \ref{\Palatini} and references quoted therein) and which is in fact the
theoretical basis to explain also the meaning of the well known alternative variables for gravity
known as {\it Ashtekar's variables} (see \ref{\DualA}, \ref{\DualC}, \ref{\DualD}, \ref{\DualE}).
The main feature of BCG theory is that it  canonically separates the contributions to conserved
quantities and entropy of gravitational and ``matter'' parts. Various contributions have been {\it
guessed} in literature (see \ref{\BTZMann}) but, as we shall show hereafter, the separation cannot
be done in BCEA theory where the various contributions are tangled up.

Let us first recall the main results of Part I which shall be useful  in the sequel.
We start from the standard Hilbert-Einstein Lagrangian with cosmological constant
$$
L=\calL\>\ds={\al}(r-2\La)\>\sqrt{g}\>\ds
\fl{\LagCCm}$$
where $\al\not=0$ is a constant
and notice that the BTZ metric (see \ref{\BTZRef}, \ref{\BTZRefB}, \ref{\BTZRefC},
\ref{\BTZRefD})
$$
\gBTZ=-N^2\dt^2+N^{-2}\dr^2+r^2(N_\phi\dt+d\phi)^2
\fl{\BHMann}$$ 
is a solution of Einstein field equations, where we set
$$
\eqalign{
&N^2=-\mu+{r^2\over l^2} + {J^2\over 4r^2} \cr
&N_\phi=-{J\over 2r^2}\cr
}\qquad\qquad\qquad
\eqalign{
&\mu={r^2_++r^2_-\over l^2} \cr
&J=2{r_+r_-\over l}\cr
}
\fn$$
Then we apply the variational prescription for N\"other theorem (see Part I)
and obtain for a generic conserved quantity $\hat Q(L,\xi,g)$ the following recipe
$$
\de_X\hat Q(L,\xi,g)=\int_{ \infty} \Big(\de_X\calU(L,\xi,g)-i_\xi \big(\F(L,g)[X]\big)
\Big)
\fl{\NewCQ}$$
where: $X$ is any deformation of configuration fields; $\de_X$ donotes variation along $X$;
$\infty$ is the asymptotic boundary of spacetime (i.e.\ space infinity);
$\xi$ is a vector field over spacetime $M$; $g$ is any Lorentzian metric over $M$;
$\calU(L,\xi,g)$ is the {\it superpotential} of the Lagrangian $\LagCCm$, given by equation $(3.7)$
in Part I;
\message{
 CONTROLLA che il superpotenziale in Part I sia formula (3.7)
}
$\F(L,g)[X]$ is the map, called {\it Poincar\'e-Cartan morphism}, given by equation $(4.9)$ in
Part I. 
\message{
 CONTROLLA che il il morfismo di PC  in Part I sia formula (4.9)
}
For conserved quantities of BTZ solutions we find 
$$
\hat Q_D(L,\del_t,\gBTZ)= 2\pi\al\> \mu
\qquad
\hat Q_D(L,\del_\phi,\gBTZ)= -2\pi\al\> J
\fl{\TotalConservedQuantities}$$
associated to symmetry generators $\del_t$ and $\del_\phi$, respectively.
By applying a suitable covariant ADM formalism one can obtain the same results by means of
the following expression
$$
\hat Q_D(L,\xi,g)=\int_{\del D} [\calU(L,\xi,g)-\tilde\calB(L,\xi,g)]
\fl{\CorrectedConsQuant}$$
Here the correction $\tilde\calB(L,\xi,g)$ is defined by pull-back along a section $g$ of the
following quantity
$$
\tilde\calB(L,\xi)=[p^{\mu\nu} w^\la_{\mu\nu}\xi^\rho+ \al\>\sqrt{h}\>
h^{\al\rho}{\mathop \na^{(h)}}{}_\al\xi^\la]\>\ds_{\la\rho}
\fl{\NewCorrectionADM}$$
where $p^{\mu\nu}={\del\calL\over\del r_{\mu\nu}}$ are the {\it covariant naive momenta},
$\ga^\la_{\mu\nu}$ denotes the Levi-Civita connection of $g$, $h=\Vert h_{\mu\nu}\Vert$ is any
reference {\it background solution}, $\Ga^\la_{\mu\nu}$ denotes the Levi-Civita connection of $h$
and we set
$$
w^\la_{\mu\nu}= u^\la_{\mu\nu}- U^\la_{\mu\nu}
\qquad\qquad
\eqalign{
&u^\la_{\mu\nu}:= \ga^\la_{\mu\nu} - \de^\la_{(\mu} \ga^\ep_{\nu)\ep}\cr
&U^\la_{\mu\nu}:= \Ga^\la_{\mu\nu} - \de^\la_{(\mu} \Ga^\ep_{\nu)\ep}\cr
}
\fn$$

We also recall that the density integrated in $\CorrectedConsQuant$ can be obtained as the
superpotential of the following first order Lagrangian
$$
\tilde L_1=[\al(r-2\La)\sqrt{g} -d_\la(p^{\mu\nu}w^\la_{\mu\nu})-\al(R-2\La)\sqrt{h}]\>\ds
\fl{\NewFirstOderLagrangian}$$
where $\d_\la$ is the formal divergence operator defined in Part I and $R$ is the scalar curvature
of the background $h$.
For what concerns entropy we recall that there is a general prescription for the variation of the
entropy
$$
\de_X \calS={1\over T}\int_\Si\Big(\de_X\calU(L,\xi,g)-i_\xi\big(\F(L,g)[X]\big)\Big)
\fl{\Entropy}$$
where $\xi=\del_t+\Om\del_\phi$, $\Om$ is the {\it angular velocity of horizon}, $T$ is the
temperature of the black hole and $\Si$ is a $(n-2)$-submanifold such that $\infty-\Si$ is a
homological boundary (i.e.\ it does not enclose any singularity). Equation $\Entropy$ is equivalent
to the first principle of thermodynamics and it gives, for any BTZ solution, the following result
$$
\calS=8\pi^2\>\al\> r_+
\fl{\UnVarEntropy}$$

We shall reproduce and compare these results in BCEA framework.

\NewSection{BCEA theories}
\CurrentSection\edef\BCEASection{\Mock}

Let us fix a  $\SO(2,1)$-principal bundle $\calF=(F,M,\pi,\SO(2,1))$ over a $3$-dimensional
spacetime manifold $M$.
Following \ref{\Bologna} we define a {\it moving frame} to be any  principal vertical morphism
$e:\calF\arr L(M)$ into the linear frame bundle of $M$, i.e.
$$
\matrix{
\hfill F & {\buildrel e\over\larr} & L(M) \hfill \cr
\hfill \pi\Big\downarrow&  & \Big\downarrow  \hfill \cr
\hfill M & {\buildrel id\over\larr} & M \hfill \cr
}
\qquad\qquad
\matrix{
\hfill F & {\buildrel e\over\larr} & L(M)\hfill  \cr
\hfill \tilde R_a\Big\downarrow &  & \Big\downarrow R_a\hfill \cr
\hfill F & {\buildrel e\over\larr} & L(M) \hfill \cr
}
\qquad
\forall a\in\SO(2,1)\subset\GL(3)
$$
where $\tilde R_a$ and $R_a$ denote the relevant canonical right actions of $a$ on principal
bundles.

Of course, depending on the choice of $M$ and $\calF$, moving frames may or may not exist (see
\ref{\Bologna}). We stress once more that by {\it (global) moving frame} we mean a global morphism
$e:F\arr L(M)$, the existence of which does not require the existence of a global section in
$L(M)$: in other words, {\it (global) moving frames} may exist even on non parallelizable
manifolds. However, if $M$ allows a metric $g$ of signature $\eta=(2,1)$ it is always possible to
choose a bundle
$\calF$ so that global moving frames exist; as in \ref{\Bologna} we call such bundles {\it
structure bundles}. In the sequel, $\calF$ will be assumed to be a structure bundle.
Although we are not directly interested here in bundle-theoretic aspects, we remark that there
exists a bundle associated to $\calF$, whose global sections corresponds to global moving frames
(see \ref{\Bologna}, \ref{\FFFG} for greater details).

Let us choose trivializations $\varsigma$ in $\calF$ and $\del_\mu$ in $L(M)$, considered as a
family of local sections as it is standard when dealing with principal fiber bundles.
Using induced fibered coordinates, one can write a moving frame as
$$
e: \varsigma \mapsto e_a\equiv e_a^\mu \del_\mu 
\fn$$
so that $e_a^\mu$ form an invertible matrix and they may be regarded as the {\it components} of the
moving frame $e$;
$e_a$ ($a=1$, $2$, $3=\dim(M)$) denote a linear frame over $M$, i.e.\ a basis in its tangent
space.
Let us consider a principal automorphism in the structure bundle $\calF$, which is locally given
by
$$
\left\{
\eqalign{
&x'=f(x)\cr
&a'=\al(x)\cdot a
}
\right.
\qquad\qquad
\eqalign{
&\al(x)\in\SO(2,1)\cr
&\cr
}
\fn$$
where $(x,a)$ are local fibered coordinates on $\calF$.
This automorphism acts on moving frames as follows
$$
(e')_a^\mu= J^\mu_\nu \> e^\nu_b\> \bar\al^b_a
\fl{\GNAction}$$
where $\Vert \bar\al^b_a \Vert$ is the inverse matrix of $\al(x)$ and $J^\mu_\nu=\del_\nu f^\mu$
is the Jacobian of the diffeomorphism $f$ induced on spacetime $M$.

Let $e^a_\mu$ denote the inverse matrix of $e_a^\mu$.
Any moving frame $e$ induces a metric $g_{\mu\nu}= e^a_\mu\>\eta_{ab}\>e^b_\nu$ where $\eta_{ab}$
is the standard diagonal matrix of signature $\eta=(2,1)$;
$g$ is defined so that its orthonormal frame bundle $\SO(M,g)$ (which is the sub-bundle of $L(M)$ of
$g$-orthonormal frames) concides with the image of the moving frame $e$.
We understand that Latin indices are raised and lowered by  $\eta_{ab}$, while Greek indices are
moved by the induced metric $g_{\mu\nu}$.
Notice that because of $e^{a\mu}=g^{\mu\nu}e^a_\nu=\eta^{ab}e_b^\mu$ no possible
confusion arises.
We shall denote by $A^{ab}_\mu$ any principal $\SO(2,1)$-connection of $\calF$ (not necessarily
induced by the frame).
In dimension three one can use the parametrization for $\SO(2,1)$-connections given by:
$$
A^a_\mu=\ep^{a\>\cdot\>\cdot}_{\>\>bc}\> A^{bc}_\mu
\fl{\OneIndex}$$
which is better suited for a comparison with the current literature (see \ref{\BTZMann}).
In this notation the  structure constants of $\SO(2,1)$ are $c^a_{\>\>bc}={1\over
2}\ep^{a\>\cdot\>\cdot}_{\>\>bc}$.

Following \ref{\BTZMann} and \ref{\CaGe}, let us also introduce the ``matter'' fields $b^a_\mu$
and
$c^a_\mu$ whose nature is analogous to
$e^a_\mu$,  though they are allowed to be degenerate.
Let us consider the $\SO(2,1)$-connection $A^{ab}_\mu$ and the Levi-Civita connection
$\ga^{\la}_{\si\mu}$ induced by the moving frame $e$ through the induced metric $g_{\mu\nu}$;
we can then define the covariant derivative of the moving co-frame $e^a_\mu$ as:
\def\naA{{\mathop \na^{_{(A)}}}}
$$
\eqalign{
&\naA_\mu e^a_\nu= \d_\mu e^a_\nu + A^{a\cdot}_{\>\>b\mu}e^b_\nu - \ga^\la_{\nu\mu}e^a_\la\cr
&\naA e^a=\naA_\mu e^a_\nu \>\>\d x^\mu\land \d x^\nu=
(\d_\mu e^a_\nu + A^{a\cdot}_{\>\>b\mu}e^b_\nu)\>\d x^\mu\land \d x^\nu\cr
}
\fn$$
and analogously for the fields $b^a_\mu$ and $c^a_\mu$.

Consider now a field theory for the configuration fields $(e_a^\mu,A^{a}_\mu,b^a_\mu,c^a_\mu)$.
Let us introduce the field strength $2$-form of the $\SO(2,1)$-connection $A^{a}_\mu$
$$
F^a={1\over 2}F^a_{\mu\nu}\>\d x^\mu\land\d x^\nu=
{1\over 2}(\d_\mu A^a_\nu-\d_\nu A^a_\mu
+ {1\over 2} \ep^{a\>\cdot\>\cdot}_{\>\>bc} A^b_\mu\> A^c_\nu)\>\d x^\mu\land\d
x^\nu
\fn$$
or equivalently of $A^{ab}_\mu$
$$
F^{ab}={1\over 2}F^{ab}_{\mu\nu}\>\d x^\mu\land\d x^\nu=
(\d_\mu A^{ab}_\nu +  A^{a\>\cdot}_{\>\>c\mu}\> A^{cb}_\nu)\>\d x^\mu\land\d x^\nu
\fn$$
One can easily obtain the relation $F^a_{\mu\nu}=\ep^{a\>\cdot\>\cdot}_{\>\>bc}\> F^{bc}_{\mu\nu}$.

We can define the BCEA Lagrangian (see \ref{\CaGe}, \ref{\BTZMann}) by:
$$
\eqalign{
L_{_{BCEA}}
=&\>\calL_{_{BCEA}}\>\ds=\>-\al\>\eta_{ab}\>\Big({1\over 2}e^a_\mu F^b_{\nu\rho}+ b^a_\mu\naA_\nu
c^b_\rho \Big)\ep^{\mu\nu\rho}\>\>\ds=\cr
=&\>\calL_{_{EA}}\>\ds+\calL_{_{BC}}\>\ds=
\>\big(-\al\>\eta_{ab}\>e^a\land F^b\big)+
\>\big(-\al\>\eta_{ab}\>b^a\land\naA c^b\big) \cr}
\fl{\BCEALagrangian}$$
This is a kind of gauge theory with $\SO(2,1)$ as structure group.
From the Lagrangian $\BCEALagrangian$ one obtains the following field equations:
$$
\eqalign{
&F^a=0\cr
&\naA e^a +{1\over 2} \ep^{a\>\cdot\>\cdot}_{\>\>bc}\> b^b\land c^c=0\cr
&\naA b^a=0\cr
&\naA c^a=0\cr
}
\fl{\BCEAfEq}$$

A possible solution for these equations, in the gauge $A^a_\mu=0$, is (see \ref{\CaGe},
\ref{\BTZMann}):
$$
e=
\left\vert\matrix{
{\la r_+\over l}&0&-\la r_-\cr
0&{l\over \nu}{\del\over \del r}({\la})&0\cr
-2{\nu r_+\over l^2}&0&2{\nu r_-\over l}\cr
}
\right\vert
\qquad\qquad\qquad\>\>
\left\{
\eqalign{
&\nu^2={r^2-r_-^2\over r_+^2-r_-^2}\cr
&\la^2={r^2-r_+^2\over r_+^2-r_-^2}\cr
}
\right.
\fl{\BCEASolutionA}$$
$$
b=
\left\vert\matrix{
{r_-\over l}&0&-r_+\cr
0&-l{\del\over\del r}(\nu+\la)&0\cr
{r_+\over l}&0&-r_-\cr
}
\right\vert
\quad
c=
\left\vert\matrix{
-{r_-\over l^2}&0&{r_+\over l}\cr
0&{\del\over\del r}(\nu-\la)&0\cr
{r_+\over l^2}&0&-{r_-\over l}\cr
}
\right\vert
\fl{\BCEASolutionB}$$
where $r_+$ and $r_-$ are the outer and inner horizons of a BTZ black hole.

Notice that the metric induced by the moving frame $\BCEASolutionA$ coincides in fact with the BTZ
solution $\BHMann$. In some sense, this particular choice for the topological matter term due to
fields $(b,c)$ {\it simulates} the negative cosmological constant of the Lagrangian $\LagCCm$.
However, we stress that these two theories are {\it not} equivalent even if for any solution
of the first theory there is a solution of the BCEA theory which induces the same metric, so that
in a suitable sense they both describe the same spacetimes (see \ref{\BTZMann}).
In BCEA theories one has indeed additional fields $(b^a,c^a)$ which, at least in principle, are
expected to contribute to mass and angular momentum. Moreover, from a mathematical viewpoint
the structure of their symmetry groups is completely different. The Lagrangian $\LagCCm$ is
covariant with respect to diffeomorphisms of spacetime, i.e.\ it describes a {\it natural theory}
owing to the fact that it is a purely metric theory. The BCEA theory is covariant with respect to
$\Aut(\calF)$; locally, this amounts to require covariance with respect to both diffeomorphisms of
spacetime and local (pseudo)-rotations. Globally, this splitting of $\Aut(\calF)$ is meaningless
since there is no natural way of defining an action of diffeomorphisms alone on a moving frame.
[Moving frames are basically a family of local linear frames whose transition functions are
$\SO(\eta)$-valued; if we let a spacetime diffeomorphism to act on the family then we still get a
family of frames but transition functions are no longer $\SO(\eta)$-valued. Thus the
transformed family of local frames do not define in general a global metric with the correct
signature on spacetime.] Having set up a geometric framework using fiber bundles together with
their automorphisms, we are able to develop a global theory which does not use the local splitting
of the symmetry group
$\Aut(\calF)$. Thus from a local viewpoint we have two (inequivalent) local theories, but from a
global viewpoint just one of them is meaningful. 

As we fixed the trivialization $\varsigma$ of $\calF=(F,M,\pi,\SO(2,1))$, any point $p\in F$
(with $\pi(p)=x\in M$) may be written as $p=R_\te \varsigma(x)$, where $\te= \|\te^a_{\>b}\|  \in
\SO(2,1)$; then we can define a pointwise basis of vertical right-invariant vector fields on
$\calF$:
$$
\si_{ab}={1\over 2}(\eta_{ac}\>\rho^c_{\>b}-\eta_{bc}\>\rho^c_{\>a})
\qquad\qquad
\rho^a_{\>b}=\te^a_{\>c}{\del\over\del \te^b_{\>c}}
\fn$$

Now, let us consider an infinitesimal symmetry generator:
$$
\Xi=\xi^\mu(x)\>\del_\mu +\xi^{ab}(x)\>\si_{ab}
\fl{\AutGen}$$
which is a right-invariant (thence projectable) vector field over the structure bundle $\calF$.
Using the action $\GNAction$ of $\Aut(\calF)$ on configuration fields we can define by standard
techniques (see \ref{\Kolar}) the Lie derivatives of fields with respect to $\Xi$ obtaining:
$$
\eqalign{
&\Lie_\Xi e^a_\mu= \naA_\nu e^a_\mu\>\xi^\nu+e^a_\nu\na_\mu\xi^\nu-e_{b\mu}\>\xi^{ab}_{_{(V)}} \cr
&\Lie_\Xi A^{a}_\mu=
F^a_{\nu\mu}\>\xi^\nu+\ep^{a\>\cdot\>\cdot}_{\>\>bc}\>\naA_\mu\xi^{bc}_{_{(V)}}\cr &\Lie_\Xi
b^a_{\mu}=\naA_\nu b^a_\mu\>\xi^\nu+b^a_\nu\na_\mu\xi^\nu-b_{b\mu}\>\xi^{ab}_{_{(V)}}\cr &\Lie_\Xi
c^a_{\mu}=\naA_\nu c^a_\mu\>\xi^\nu+c^a_\nu\na_\mu\xi^\nu-b_{b\mu}\>\xi^{ab}_{_{(V)}}\cr }
\fn$$
where we have set $\xi^{ab}_{_{(V)}}=\xi^{ab}+ A^{ab}_\mu\xi^\mu$ for the vertical part of $\Xi$
with respect to the connection $A^{ab}_\mu$ (see \ref{\FFFG}).

By considering the variation of the Lagrangian $\BCEALagrangian$ and integrating by parts, we
obtain a term related to field equations $\BCEAfEq$ plus the divergence of the following
$2$-form:
$$
\eqalign{
\F(L_{_{BCEA}},g)[\de A,\de c]=&\F_{_{EA}}[\de A]+\F_{_{BC}}[\de c]=\cr
=&\big(\al\>\eta_{ab}\> b^a\land \de c^b \big)+ \big(\al\>\eta_{ab}\> e^a\land \de A^b \big)
\cr}
\fl{\BCEAF}$$
which is the {\it Poincar\'e-Cartan morphism} for the BCEA theory.

Then, owning to $\Aut(\calF)$-covariance and  N\"other's
theorem, the Lagrangian $\BCEALagrangian$ admits the following superpotential $1$-form
$$
\eqalign{
\calU(L_{_{BCEA}},\Xi)=&\calU_{_{EA}}[\Xi]+\calU_{_{BC}}[\Xi]=\cr
=&\big(\al\> \xi^{ab}_{_{(V)}}\>\ep_{abc}\> e^c\big)+
\big(-\al\> \xi^\nu\> c_{a\nu}\> b^a  \big) 
\cr}
\fn$$
where $\Xi$ is given by $\AutGen$.
Thus, we can define the variation of conserved quantities as
$$
\de\hat Q_D(L_{_{BCEA}},\Xi,g)=\int_{\del D}
\Big(\de\calU(L_{_{BCEA}},\Xi,g)-i_\xi\F(L_{_{BCEA}},g)[\de A,\de c]\Big)
\fl{\VCQ}$$
where $\xi=\xi^\mu\del_\mu$ is the projection of the symmetry generator $\Xi$ on spacetime $M$.

We again remark that the BCEA theory is covariant with respect to $Aut(\calF)$
and there is no natural action of  spacetime diffeomorphisms on configuration fields.
[This is basically due to the fact that the local action of diffeomorphisms and pure gauge
transformations do not commute.]
Thence in previous formulae the infinitesimal generators $\xi^\mu$ and
$\xi^{ab}$ are so far completely unrelated.
If we want to compare conserved quantities along the solution $\BCEASolutionA$ and $\BCEASolutionB$
with purely metric theories, which are $\Diff(M)$-covariant,
we need to establish some relation among infinitesimal symmetry generators.

\NewSection{Kosmann lift}

As previously remarked (see also \ref{\FFFG} and references quoted therein), despite there is no
natural action of
$\Diff(M)$ on the configuration bundle, one can define a global action of {\it infinitesimal}
symmetry generators. In geometrically oriented literature this is known as the {\it Kosmann lift of
vector fields}. Basically, one starts with a vector field $\xi=\xi^\mu\del_\mu$ on spacetime $M$
and lifts it by means of the natural lift to the frame bundle $L(M)$
defined by
$$
\hat\xi=\xi^\mu(x)\>\del_\mu+\d_\mu\xi^\nu(x)\>\rho^\mu_\nu
\qquad\qquad
\rho^\mu_\nu=V^\mu_a{\del\over\del V^\nu_a}
\fl{\naturalLift}$$
where $(x^\mu,V^\mu_a)$ are fibered coordinates on $L(M)$ so that $\rho^\mu_\nu$ is a pointwise
basis for vertical right-invariant vector fields.

Now, by fixing a moving frame $e:\calF\arr L(M)$ one would somehow pull-back $\hat\xi$
over $\calF$.
Unfortunately, the pull-back of vector fields can be performed only along bijective maps, while
here $e$, though injective, is not surjective.
Nevertheless, if $\hat\xi$ were tangent to the image of $\calF$ through the moving frame $e$,
its pull-back would be well defined.
Thus, if we define a global way of projecting $\hat\xi$ to the tangent space of
$\Im(e)$ we can use it to define a global (though non natural) lift of vector fields over $M$
to vector fields over $\calF$. This lift depends on the moving frame $e$ we chose
and it establishes a relation between the generators $\xi^\mu$ and $\xi^{ab}$ of infinitesimal
symmetries.

Once we choose a moving frame $e$, the $\SO(2,1)$-connection $\om^{ab}_\mu$, called the {\it spin
connection}, is induced. It is {\it compatible} with the frame, i.e.
$$
{\mathop \na^{(\om)}}_\mu e_a^\nu=\d_\mu e_a^\nu +\ga^\nu_{\la\mu}\>e_a^\la
-\om^{b\>\cdot}_{\>\>a\mu}\>e_b^\nu=0
\fn$$
or, equivalently, it is related to the Levi-Civita connection $\ga^\la_{\mu\nu}$ of the metric
$g_{\mu\nu}$ induced by the frame $e_a^\mu$ itself as follows:
$$
\om^{ab}_\rho=e^a_\la\big( \ga^\la_{\mu\rho}\>e^{b\mu}+\d_\rho e^{b\la}\big)
\fn$$
Let $(x^\mu,\te^a_{\>b})$ be the fibered coordinates induced in $L(M)$ by the trivialization
$\varsigma$ fixed on $\calF$ and the moving frame $e$; these {\it adapted coordinates} induce a
pointwise basis for vertical right invariant vector fields:
$$
\rho^a_{\>b}=\te^{a}_{\>c}{\del\over\del \te^b_{\>c}}
\fn$$ 
We can recast the natural lift $\naturalLift$ with respect to these adapted coordinates as
$\hat\xi=\xi^\mu(x)\>\del_\mu+\hat\xi^{a}_{\>b}\>\rho^b_{\>a}$ by setting
$$
\hat\xi^{a}_{\>b}=
e^{\mu}_b\>{\mathop \na^{(\om)}}_\mu (e^a_\nu\xi^\nu)-\om^{a\>\cdot}_{\>\>b\mu}\xi^\mu=
e^a_\rho\big(e_b^\mu\>\d_\mu\xi^\rho- \xi^\mu \>\d_\mu e_b^\rho\big)
\fl{\KA}$$

We can now define the Kosmann lift (see \ref{\Brno}) as
$$
\hat\xi_{_{(K)}}=\xi^\mu(x)\>\del_\mu+\hat\xi^{ab}(x)\>\si_{ab}
\qquad
\hat\xi^{ab}=\hat\xi^{[a}_{\>\>c}\eta^{b]c}
\qquad
\si_{ab}=\eta_{c[a}\rho^{c}_{\>\>b]}
\fl{\KosmannLift}$$
We remark that $\si_{ab}$ is a pointwise basis for vertical right-invariant vector fields on the
sub-bundle $\Im(e)$; $\si_{ab}$ also denote the induced vector fields on $\calF$.
The projection of vector fields of $L(M)$ over $\Im(e)$ is thence made by skew-symmetrization.
We stress that the components $\hat\xi^{ab}=\hat\xi^{[a}_{\>\>c}\eta^{b]c}$ of the Kosmann lift do
not depend on any connection, as clearly shown by equation $\KA$.

We also stress that in BCEA theories the Kosmann lift $\KosmannLift$ is just one of
various possibilities of establishing a relation among $\xi^\mu$ and $\xi^{ab}$.
The relation among the infinitesimal symmetry generators $\xi^\mu$ and $\xi^{ab}$
certainly requires a {\it global} lift, which possibly does not depend on fields
other than configuration ones in order to preserve covariance.
However, in BCEA theory there are two $\SO(2,1)$-connections (namely, $\om^{ab}_\mu$ and
$A^{ab}_\mu$) which thence enable us to define (at least) four well-defined global lifts, given by
the following expressions:
$$
\eqalignno{
&\hat\xi^{a}_{\>b}= e^{\mu}_b\>{\mathop \na^{(\om)}}_\mu
    (e^a_\nu\xi^\nu)-\om^{a\>\cdot}_{\>b\mu}\xi^\mu  &\Getfl{\KLiftA}  \cr
&\hat\xi^{a}_{\>b}=e^{\mu}_b\>{\mathop \na^{(A)}}_\mu
    (e^a_\nu\xi^\nu)-\om^{a\>\cdot}_{\>b\mu}\xi^\mu  &\Getfl{\KLiftC}  \cr
&\hat\xi^{a}_{\>b}=e^{\mu}_b\>{\mathop \na^{(\om)}}_\mu
    (e^a_\nu\xi^\nu)-A^{a\>\cdot}_{\>b\mu}\xi^\mu  &\Getfl{\KLiftB}  \cr 
&\hat\xi^{a}_{\>b}=e^{\mu}_b\>{\mathop \na^{(A)}}_\mu
    (e^a_\nu\xi^\nu)-A^{a\>\cdot}_{\>b\mu}\xi^\mu &\Getfl{\KLiftD}  \cr
}
$$
We stress again that all these lifts are global (though not natural) and in principle no one of
them is, {\it a priori} , better than the others.

\NewSection{Conserved quantities and entropy in BCEA theories}

We are now in the position of calculating the variation of the conserved quantities $\VCQ$
along the solution $\BCEASolutionA$ and $\BCEASolutionB$ associated to the diffeomorphism generators
$\del_t$ and $\del_\phi$ through the lifts $\KLiftA$-$\KLiftD$.
The contribution of $\F$ identically vanishes, i.e.:
$$
\int_{\del D} i_{\del_t}\F(L_{_{BCEA}},g)[\de A,\de c]=0\>,
\qquad
\int_{\del D} i_{\del_\phi}\F(L_{_{BCEA}},g)[\de A,\de c]=0
\fn$$
We remark that this result does not rely on the choice of the lift because the vertical part of
the symmetry generator does not enter expression $\BCEAF$.
Let us recall that the vertical part of a lifted symmetry generator $\hat\xi$ is
$$
\xi^{ab}_{_{(V)}}\si_{ab}=\big(\hat\xi^{ab}+A^{ab}_\mu\xi^\mu\big)\si_{ab}
\fl{\XiVertical}$$ 
where $\hat\xi^{ab}$ may be any one of the lifts defined by $\KLiftA$-$\KLiftD$;
we thence obtain respectively the following results
$$
\eqalignno{
& \de \hat Q[\del_t]=\al\>\hbox{${2\pi\over l}$}\de J
   \qquad\qquad
   \de \hat Q[\del_\phi]=-2\pi\>\al\> l\>\de \mu
    &\Getfl{\DQA}  \cr
& \de \hat Q[\del_t]=\al\>\hbox{${2\pi\over l}$}\de J
   \qquad\qquad
   \de \hat Q[\del_\phi]=2\pi\>\al\> (\hbox{${1\over 2}$}\>\de J-l\>\de \mu)
    &\Getfl{\DQB} \cr
& \de \hat Q[\del_t]=\al\>\hbox{${2\pi\over l}$}\de J
   \qquad\qquad
   \de \hat Q[\del_\phi]=-2\pi\>\al\> (\de J+l\>\de \mu)
   &\Getfl{\DQC}  \cr
& \de \hat Q[\del_t]=\al\>\hbox{${2\pi\over l}$}\de J
   \qquad\qquad
   \de \hat Q[\del_\phi]=-2\pi\>\al\> (\hbox{${1\over 2}$}\>\de J+l\>\de \mu)
   &\Getfl{\DQD}  \cr
}
$$
where integrals are performed on $1$-spheres $S^1_r$ of radius $r$.
One can easily integrate these quantities to obtain $\hat Q[\del_t]$ and $\hat Q[\del_\phi]$.
We stress that all these quantities are conserved, though of course at most one of them
can be interpreted as the {\it mass} and the other as the {\it angular momentum}.
Notice that the quantities $\DQA$ agree with the result found in \ref{\BTZMann} using a locally
defined (and globally ill-behaved) action of spacetime diffeomorphisms on configuration fields;
notice in particular the exchange between the mass and the angular momentum:
here they are related to the original (global) Kosmann lift $\KosmannLift$.
In other words, angular momentum $\mu$ is not always generated by $\del_\phi$ but rather by another
symmetry generator
$\la'$ given  respectively by
$$
\eqalignno{
&\la'=\del_\phi
    &\Getfn  \cr
& \la'=\del_\phi -{l\over 2}\del_t
    &\Getfn \cr
& \la'=\del_\phi+l\del_t
   &\Getfn  \cr
& \la'=\del_\phi+{l\over 2}\del_t
   &\Getfn  \cr
}
$$

As already stressed in \ref{\BTZMann}, the total conserved quantities  determined here differ from
the total conserved quantities $\TotalConservedQuantities$ determined for the metric
theory $\LagCCm$.
As we already noticed, despite both theories describe the same spacetimes, they are clearly
dynamically inequivalent, a fact that reverberates on the difference of conserved quantities. 
Finally, as shown above, one could directly compute conserved quantities by fixing a background as
in 
$\CorrectedConsQuant$ or, equivalently, by using a first order Lagrangian
analogous to $\NewFirstOderLagrangian$, obtaining the same results.
Again the background is generally needed to avoid divergences.

Now, we can consider the Killing vector $\xi=\del_t+\Om\>\del_\phi$ to calculate the entropy
$\Entropy$
 using the Kosmann lift $\KosmannLift$; we obtain
$$
\de_X\calS=8\pi^2\>\al\>\de r_-
\fl{\BCEAEntropy}$$
which clearly gives $\calS=8\pi^2\>\al\> r_-$.
Notice the exchange of the role of the inner and outer horizons in the expression of entropy if
compared with $\UnVarEntropy$.
Once again a different result with respect to
$\UnVarEntropy$  is due to the inequivalence of the theories.

The entropy has been calculated
in a geometric framework; this result corroborates the results already obtained in \ref{\BTZMann}.
We remark that if one tries to use one of the other lifts $\KLiftC$, $\KLiftB$ or $\KLiftD$,
the expression for the variation of the entropy so obtained turns out to be non-integrable.
Similarly, non-integrable expressions are obtained using the spin connection $\om^{ab}_\mu$
instead of $A^{ab}_\mu$ to define the vertical projection $\XiVertical$, i.e.\
$\xi^{ab}_{_{(V)}}=\hat\xi^{ab}+\om^{ab}_\mu\xi^\mu$.
It once again selects the original Kosmann lift as a kind of  {\it canonical choice}; a further
hint in favour of this choice will arise in next Section.

Finally, we remark that another possibility for the lift of vector fields over $M$ to vector
fields over $\calF$ is to use directly the dynamical connection $A^{ab}_\mu$, as one does in gauge
theories. In this case one puts
$$
\xi^{ab}=-A^{ab}_\mu\xi^\mu
\fl{\GaugeLift}$$
which obviously gives $\xi^{ab}_{_{(V)}}=0$.
It can be easily shown that by applying this choice to our example one obtains the same conserved
quantities as in
$\DQA$ and thence the same entropy. Curiously enough, these two lifts are therefore completely
equivalent in the example under investigation, though they are completely unrelated in general.
A comparison between them should then  be carried over in other examples.

\NewSection{Purely metric BCG theories}

As we said, in BCEA theory two $\SO(2,1)$-connections are involved, namely the
dynamical connection $A^{ab}_\mu$ and the spin connection $\om^{ab}_\mu$; as we noticed in
the previous Sections, the freedom we have in choosing between them causes ambiguities when dealing
with conserved quantities.
 
To overcome these difficulties, we hereafter want to define a field theory (BCG) equivalent
(on-shell) to BCEA theory introduced in Section \BCEASection, in which only tensorial objects
appear. Through a partial (covariant) Legendre transformation the BCEA Lagrangian $\BCEALagrangian$
will be rewritten into a purely metric form in which the gravitational field is described by means
of a metric on spacetime in place of a $\SO(2,1)$-connection together with a moving frame as it
happens in BCEA theory.

In BCEA theory the connection $\om^{ab}_\mu$ is determined by the compatibility condition
$$
{\mathop \na^{(\om)}} e^a =0
\fl{\OmegaConstrain}$$
while the connection $A^{ab}_\mu$ satisfies field equation $\BCEAfEq$:
$$
{\mathop \na^{(A)}} e^a +{1\over 2} \ep^{a\>\cdot\>\cdot}_{\>\>bc}\> b^b\land c^c=0
\fl{\AConstrain}$$
If the ``matter'' Lagrangian $L_{_{BC}}=\calL_{_{BC}}\>\ds$ were not present (or more generally if
it did not depend on the connection)
then the field equation $\AConstrain$ would reduce to a compatibility condition for $A^{ab}_\mu$
which ensures that $A^{ab}_\mu=\om^{ab}_\mu$ on-shell.
When this is the case, according to the so-called Palatini method (also known as {\it first order
variational method}, see \ref{\Palatini} and references quoted therein), one would calculate the
connection with respect to the moving frame
$e$ and its first order derivatives. Then by substituting the expression so obtained into the
Lagrangian one can recast it into the standard Hilbert-Einstein second order Lagrangian
(possibly coupled with matter fields).

The ``matter'' Lagrangian 
$L_{_{BC}}= -\al\>\eta_{ab}\>b^a_\mu\naA_\nu c^b_\rho \>\ep^{\mu\nu\rho}\>\>\ds$
actually depends on the connection $A^{ab}_\mu$ through the covariant derivative $\naA_\nu
c^b_\rho$. Since $A^{ab}_\mu$ has to satisfy equation $\AConstrain$, it is no longer equal to the
(compatible) spin connection $\om^{ab}_\mu$.
Nevertheless, since the matrix $e^a_\mu$ associated to the moving (co)frame is invertible,
equation $\AConstrain$ can anyway be solved with respect to the equivalent connection $A^{a}_\mu$
(see $\OneIndex$), giving a (unique) function of the moving frame $e$, its first order
derivatives, together with
$b$ and
$c$ fields. By an easy (though tedious) calculation we obtain in fact 
$$
A^{a}_\mu= \ep^{a\>\cdot\>\cdot}_{\>\>bc}\>\om^{bc}_\mu+ 
e^a_\ga\Big[
\ep^{\ga\si\nu}\ep_{\rho\la\mu}\> b^\rho_\si \> c^\la_\nu +{1\over 2}\de^\ga_\mu\>
\Big(\>\Tr(b\cdot c)- \Tr(b)\> \Tr(c) \Big)
\Big]
\fl{\AdiOmega}$$
where $\Tr$ denotes the trace of matrices and  where we set
$$
b^\rho_\la= \>e^\rho_a \>b^a_\la
\qquad\qquad
c^\rho_\la= \>e^\rho_a \>c^a_\la
\fn$$
The pull-back of the Lagrangian $\BCEALagrangian$ through the equation $\AdiOmega$ (roughly
speaking, substituting $\AdiOmega$ into the Lagrangian $\BCEALagrangian$) defines the purely metric
BCG Lagrangian
$$
L_{_{BCG}}=L_{_{\hbox{\sevenrm H}}}+L_{_{\hbox{\sevenrm Int}}}+ L_{_{\hbox{\sevenrm Div}}}+L_{_{0}}
\fl{\BCGLagrangian}$$
where we set
$$
\eqalignno{
&L_{_{\hbox{\sevenrm H}}}=\al\>\sqrt{g}\> r \>\ds&\Getfl{\LHilbert}\cr
&L_{_{\hbox{\sevenrm Int}}}= -\al\> g_{\ga\la}\>\ep^{\mu\nu\rho}\> b^\ga_\mu \>\na_\nu c^\la_\rho
\>\ds
   &\Getfl{\LInt}\cr
&L_{_{\hbox{\sevenrm Div}}}= \al\>\d_\nu \Big[ g_{\ga\la}\>\ep^{\ga\rho\si}\Big(
b^\nu_\rho\> c^\la_\si - b^\la_\rho\> c^\nu_\si\Big)\Big]\>\ds
   &\Getfl{\LDiv}\cr
&L_{_{0}}={\al\>\sqrt{g}\over 8} \Big[ 2\Big( \Tr(b^2)\>\Tr(c^2)-2\>\Tr(b^2\cdot
c^2)+\Tr(b\cdot c)^2\Big)+\cr
&\quad\qquad\qquad\>\>-\Big(\Tr(b\cdot c) -\Tr(b)\>\Tr(c)\Big)^2\Big]\>\ds
   &\Getfl{\LZero}\cr
}
$$
where, from now on, $\na$ denotes the (metric) covariant derivative with respect to the Levi-Civita
connection of the metric $g$.  
The BCG Lagrangian is (on-shell) {\it equivalent} to BCEA Lagrangian $\BCEALagrangian$: one
obtains, in fact, field equations of the BCG Lagrangian by substituting the expression $\AdiOmega$
into the field  equations $\BCEAfEq$ of the BCEA Lagrangian. 

Geometrically, one can interpret the expression $\AdiOmega$ as the definition of the
{\it partial (covariant) dual Legendre map}
$$
(e^a_\mu, \d_\nu e^a_\mu; b^a_\mu, c^a_\mu)\mapsto
(e^a_\mu, A^a_\mu(e, \d e; b,c); b^a_\mu, c^a_\mu)
\fl{\DualLegiandreTransf}$$ 
In fact, following the multisymplectic approach to the Hamiltonian formulation of field theories
(see
\ref{\DualA}, \ref{\DualE}, \ref{\DualF} and references quoted therein) we consider $A^a_\mu$ as
the configuration fields,  while their {\it covariant momenta} are not the derivatives of the
Lagrangian density with respect to time derivatives of the
$A^a_\mu$ (as in standard $(3+1)$ Hamiltonian formulation) but rather the derivatives of
the Lagrangian density with respect to {\it all} {\it ``generalized velocities''} $F^a_{\mu\nu}$,
namely
$$
p_a^{\mu\nu}={\del \calL_{_{BCEA}}\over \del F^a_{\mu\nu}}=-{\al\over2} e_{a\rho}\ep^{\rho\mu\nu}
\fn$$
Following this approach, the moving frames $e$ are, in practice, the conjugate fields to
$A^a_\mu$.  
In this way, the BCEA Lagrangian may be considered as the analogous of the Helmholtz Lagrangian
$L_{_{H}}=p\>\dot q- H(p,q)$ in Classical Mechanics, where the mechanical analogy is established by
the correspondence $e\arr p$, $A\arr q$ and $F\arr \dot q$;
the term $p\>\dot q$ in $L_{_{H}}$ corresponds to the part $L_{_{EA}}=-\al\>\eta_{ab}\>e^a\land
F^b$ of the BCEA Lagrangian and $H$ to the part $L_{_{BC}}$.

In Mechanics, provided the regularity condition
$\det\vert{\del^2 H\over \del p\del p}\vert\not= 0$ is satisfied, 
one can (at least locally) solve the Hamilton equation
$$
\dot q={\del H\over \del p}(q,p)
\fl{\firstHamiltonEq}$$
with respect to $p$ by expressing it as a function $\tilde p(q,\dot q)$.
This defines the Legendre transformation
$$
(q, \dot q)\arr \big(q, \tilde p(q,\dot q) \big) 
\fl{\MechLegendreTransf}$$
which can be used to define the Lagrangian $L(q,\dot q)= \tilde p\>\dot q -H(q,\tilde p)$.
[This approach may be used to give, for example, purely affine formulations of General Relativity
(see \ref{\DualA}, \ref{\DualB}, \ref{\DualC}, \ref{\DualD}, \ref{\DualE}). As far as
BCEA theory is concerned, however, one cannot follow this route since the Hamilton equation
$\firstHamiltonEq$ corresponds to the field equation
$F^a=0$ which cannot be solved with respect to $e$ since it does not appear at all in the
relevant expression!]

On the contrary, one can consider the other Hamilton equation
$$
\dot p=-{\del H\over \del q}(q,p)
\fl{\pPunto}$$
Provided the regularity condition $\det\vert{\del^2 H\over \del q\del q}\vert\not= 0$ is
satisfied, the equation $\pPunto$ may be solved with respect to $q$ by expressing it as a
function
$\tilde q(p,\dot p)$. It then defines the {\it dual Legendre transformation}
$$
(p, \dot p)\arr \big(p, \tilde q(p,\dot p) \big) 
\fl{\MechDualLegendreTransf}$$
which is analogous to $\DualLegiandreTransf$, being the Hamilton equation $\pPunto$ analogous to
field equation $\AConstrain$.
Then it has been shown in general (see \ref{\DualA}) that the other set of Hamilton equations
$\firstHamiltonEq$, once we substitute $\tilde q$, may be viewed as the Lagrange equations of a
{\it dual Lagrangian}
$L^\ast(p,\dot p, \ddot p)$ which is linear in $\ddot p$.
This dual Lagrangian is the mechanical analogous of the BCG Lagrangian we built (which is clearly
linear in second order derivatives of the metric $g_{\mu\nu}$). 

Now, let us calculate conserved quantities.
First of all, we remark that, for any Lagrangian $L$, the quantity
$\de_X\calU(L,\xi,g)-i_\xi\F(L,g)[X]$ is invariant with respect to the addition of total
divergences to the Lagrangian. Thence it is not at all affected by the term $\LDiv$ in the BCG
Lagrangian. Secondly, being the term $\LZero$ a zeroth order Lagrangian, it affects field
equations but it is irrelevant to both $\calU(L_{_{BCG}},\xi,g)$ and $\F(L_{_{BCG}},g)$.
Thus, despite the BCG Lagrangian $\BCGLagrangian$ is quite complicated, the only terms in the
Lagrangian which are relevant to conserved quantities are
$L_{_{\hbox{\sevenrm H}}}$ and $L_{_{\hbox{\sevenrm Int}}}$.

By easy calculations, one can find the following
$$
\eqalign{
&\calU(L_{_{\hbox{\sevenrm H}}},\xi)=\al\> \sqrt{g}\> \na^\mu\xi^\nu\>\ep_{\nu\mu\la}\>\d x^\la\cr
&\calU(L_{_{\hbox{\sevenrm Int}}},\xi)=\al\>\xi^\ga\>\Big[
\Tr(c) b^{\>\cdot}_{\ga\la} 
+\Tr(b) c^{\>\cdot}_{\ga\la}
-b^{\>\cdot}_{\ga\nu} c^\nu_{\>\la}
-c^{\>\cdot}_{\ga\nu} b^\nu_{\>\la}
-b^{\>\cdot}_{\nu\la} c^\nu_{\>\ga}+\cr
&\qquad\qquad\qquad\qquad
+{1\over 2}g_{\la\ga}\Big(
\Tr(b\cdot c)-\Tr(b) \Tr(c)
\Big)
\Big]\>\d x^\la\cr
&\F(L_{_{\hbox{\sevenrm H}}},g)[\de g]=\al\>(g^{\la\rho}g_{\mu\nu}-\de^\la_{(\mu}\de^\rho_{\nu)})
\na_\rho \de g^{\mu\nu}\>\sqrt{g}\>\ds_\la\cr
&\F(L_{_{\hbox{\sevenrm Int}}},g)[\de g, \de c]= \al\>\Big[\De^{\mu\ga\nu}\>\de g_{\ga\nu}
- \ep^{\mu\rho\si}\> b^{\>\cdot}_{\la\si}\>\de c^\la_{\>\rho}\Big]\>\ds_\mu\cr
}
\fn$$
where we set
$$
\De^{\mu\ga\nu}={1\over 2}\Big[
\ep^{\rho\nu\si}\> b^\mu_{\>\rho}\> c^\ga_{\>\si}
-\ep^{\rho\mu\si}\> b^\ga_{\>\rho}\> c^\nu_{\>\si}
-\ep^{\rho\ga\si}\> b^\nu_{\>\rho}\> c^\mu_{\>\si}
\Big]
\fn$$
Replacing these expressions back into expression $\NewCQ$ we obtain the following variations of
conserved quantities for BTZ solutions
$$
\eqalign{
&\de \hat Q(L_{_{BCG}},\del_t,g)= \al\> {4\pi\over l^2}\>\Big[ r_-\>\de r_+\>+r_+\>\de r_-\Big]\cr
&\de \hat Q(L_{_{BCG}},\del_\phi,g)= -\al\>{4\pi\over l}\> \Big[ r_+\>\de r_+\>+r_-\>\de
r_-\Big]\cr }
\fn$$
which are readily integrated to
$$
\hat Q(L_{_{BCG}},\del_t,g)= \al\> {2\pi\over l}J
\qquad\qquad
\hat Q(L_{_{BCG}},\del_\phi,g)= -2\pi\>\al\>l\> \mu
\fl{\BCGConservedQuantities}$$
[We stress once again that the same result would be obtained by choosing as a background another
BTZ solution with different parameters $(r^{_0}_+,r^{_0}_-)$ and then using the correction
$\NewCorrectionADM$.]

As a consequence, the variation of the entropy (see eq. $\Entropy$) has the value
$\de\calS= 8\pi^2 \>\al\> \de r_-$ so that $\calS$ is given by:
$$
\calS=8\>\pi^2\>\al\> r_-
\fl{\BCGEntropy}$$

We remark that the BCG theory is a natural theory since any dynamical field is a tensor over
spacetime. Naturality means that there is a canonical action of spacetime diffeomorphisms
over dynamical fields and no ambiguity arise about the lift to be used in N\"other theorem:
a unique natural lift exists.
Such a natural lift reproduces the results $\DQA$ and $\BCEAEntropy$ and this again justifies {\it
a posteriori} our choice of the Kosmann lift made in previous Sections.

We also remark that here the {\it minimal coupling} is physically meaningful, i.e.\ we can split
the Lagrangian into a purely metric Lagrangian
$L_{_{\hbox{\sevenrm H}}}$, namely the Hilbert-Einstein Lagrangian, and ``matter'' and interaction
Lagrangian
$L_{_{\hbox{\sevenrm Int}}}+ L_{_{\hbox{\sevenrm Div}}}+L_{_{0}}$ not containig second
derivatives of the metric field. The conserved quantities $\BCGConservedQuantities$ and the entropy
$\BCGEntropy$ receive thence two contributions, each one from each term of the splitting. These are
$$
\eqalign{
\hat Q_{g}\equiv &\hat Q(L_{_{\hbox{\sevenrm H}}},\del_t,g)= 2\pi\>\al\>\mu\cr
& \hat Q(L_{_{\hbox{\sevenrm H}}},\del_\phi,g)=-2\pi\>\al\>J\cr
&\calS_g=8\>\pi^2\>\al\> r_+\cr
}
\qquad\qquad
\eqalign{
\hat Q_{_{BC}}\equiv &\hat Q(L_{_{\hbox{\sevenrm Int}}},\del_t,g)=2\pi\>\al\>\big(\hbox{$1\over
l$}J -  \mu\big)\cr
& \hat Q(L_{_{\hbox{\sevenrm Int}}},\del_\phi,g)=2\pi\>\al\>\big(J + l
\mu\big)\cr
&\calS_{_{BC}}=8\>\pi^2\>\al\>( r_--r_+)\cr
}
\fn$$
These results thoroughly justify the suggested  interpretation of the  exchange
of inner and outer horizon in $\BCEAEntropy$ as the resulting contribution of $b^a$ and $c^a$
fields to total entropy (see \ref{\BTZMann}).
They also show how the same idea cannot be applied to BCEA theory where, because of
the expression $\AdiOmega$, it is impossible to isolate in the BCEA Lagrangian a purely
gravitational part analogous to $L_{_{\hbox{\sevenrm H}}}$.
Roughly speaking, the Lagrangian $L_{_{EA}}$, when rewritten as a metric Lagrangian by using
$\AdiOmega$, does not correspond to the Hilbert-Einstein vacuum Lagrangian $L_{_{\hbox{\sevenrm
H}}}$ in $\BCGLagrangian$.

\NewSection{Conclusion and Perspectives}

We introduced and analysed BCEA and BCG theories.
Conserved quantities have been calculated for BCEA theories but, since there is an additional gauge
invariance, additional conserved quantities have arisen. They are related to pure gauge
transformations (i.e.\ local (pseudo)-rotations).
Moreover, since the symmetry group of the theory does not naturally split into a vertical part
(related to pure gauge transformations) and a horizontal part (related to diffeomorphisms of
spacetime), the problem of a global definition of mass and angular momentum arises.
We have faced this problem by considering various lifts which, although not canonical, are still
globally well-defined.
The so-called Kosmann lift was introduced in a different context (see \ref{\Brno}) to justify a
proposal due to Kosmann for the Lie derivative of spinor fields with respect to general (i.e.\ not
necessarily Killing) vector fields tangent to spacetime.
The generalizations of the Kosmann lift here considered were suggested by the fact that in BCEA
theory there are two dynamical connections, namely $A^{ab}_\mu$ and $\om^{ab}_\mu$.

The lifts we considered usually produce different quantities, which, even if conserved, are not
directly the expected {\it mass} and {\it angular momentum} of a BTZ black hole solution, but a
combination of the two, owing to a pure gauge contribution. Two of them, i.e.\ the Kosmann lift
and a suitably defined {\it gauge lift} produced conserved quantities which both agree with values
predicted by BCG theory, where there is no ambiguity  in the definition of conserved quantities,
being the latter a natural (generally covariant) theory. This result suggests that these two lifts
have to be somehow preferred with respect to the other possible lifts considered.
This claim is also supported by the fact that the equation  obtained for the variation of the
entropy is integrable only for these two lifts, and the entropy computed again agrees with the BCG
result.
However, we stress that both lifts (the Kosmann and the gauge one) are well-defined: they do not
depend on non-dynamical background fields, so that they are completely determined in terms of the
solution under consideration, and moreover they both produce the same conserved quantities.
As far as we know there is no reason, other than aesthetical, to choose between them when
considering a BTZ solution.
Thence future investigations will be addressed to study other solutions in which the two lifts
might give different predictions which could possibly help in selecting a {\it better} one from
good standing physical and/or mathematical grounds.

The BCG theory is defined starting from BCEA theory and performing a dual covariant Legendre
transformation.
We stress that the so-called Palatini (or first order) variational method (see \ref{\Palatini}) is
a particular case of the Legendre transformation we performed.
Palatini method correctly applies to frame-affine theories when the coupling to matter fields does
not depend on derivatives of the moving frame $e$.
Using the mechanical analogy we introduced, one could say that, if a Lagrangian
$L={m\over 2}(\dot q)^2-U(q)$ is considered, Palatini method  corresponds to
define momenta as $p=m\>\dot q$ while for more general Lagrangians
$L={m\over 2}(\dot q)^2-U(q,\dot q)$ a contribution is due to the generalized potential,
i.e.\ $p={\del L\over \del \dot q}=m\>\dot q-{\del U\over \del \dot q}$.
In BCEA Lagrangian, in fact, the coupling is defined through covariant derivatives, so that it
depends on the dynamical connection $A^a_\mu$: an extra contribution $b\land c$ arises so that
the dynamical connection $A^{ab}_\mu$ does not coincide with the spin connection
$\om^{ab}_\mu$.

Furthermore, extra degrees of freedom owned by the moving frame (i.e.\ those related to local
rotations) desappear when performing the Legendre transformation.
This is an effect of the covariant choice of momenta. 
Thence BCG theory is a natural theory, i.e.\ it is generally covariant with respect to
diffeomorphisms of spacetime.
Contrarily to BCEA theory, in this theory there is a well defined action of diffeomorphisms on
dynamical fields, due to their tensorial character.
Derivation of mass and angular momentum is thence a standard application of a general framework
(see \ref{\Remarks} and references quoted therein) and no ambiguities are involved.
Thus, one could say that BCG theory provided the expected values that BCEA theory should
reproduce by constraining the choice of the lift. 

Future investigations will be devoted to study the general mechanism which is behind the loss
of naturality when passing from a purely metric to a frame-affine (or purely affine) formalism
(see \ref{\DualA}, \ref{\XI}, \ref{\XII}, \ref{\XVI}, \ref{\XIV}).
In these cases some sort of lift has always to be defined to replace the natural lift one has in
the natural purely metric formalism, though no general prescription is known to be {\it a priori}
correct.   
Furthermore, the BTZ solution has provided us a good test model for studying the
transformation laws of conserved quantities (and entropy) under Legendre transformations.
Other examples have been already investigated in literature (see \ref{\Maeda} and references
quoted therein) in a different framework. We claim that this subject deserves a
geometrical setting in a more general theoretic context, which is currently under consideration
and will form the subject of forthcoming papers.

\hs
\NewSection{Acknowledgments}

We are grateful to R.\ Mann for having addressed our attention to the BTZ solution and for useful
discussions about it.

\NewSection{References}

\Biblio

\end